\documentclass[amssymb,prd,superscriptaddress,twocolumn,aps,nofootinbib,showpacs]{revtex4-1}
\usepackage{graphicx, epsfig, amssymb} 
\usepackage{amsmath, amsfonts}
\usepackage{bm} 
\usepackage[breaklinks]{hyperref}
\usepackage{color}

\def\be{\begin{equation}}
\def\ee{\end{equation}}
\def\beq{\begin{eqnarray}}
\def\eeq{\end{eqnarray}}

\begin{document}

\title{Giant Pulsar Glitches and the Inertia of Neutron-Star Crusts}

\author{T. Delsate}
\affiliation{Theoretical and Mathematical Physics Dept., Universit\'e de Mons, 20, 
Place du Parc, 7000 Mons, Belgium}
\author{N. Chamel}
\affiliation{Institut d'Astronomie et d'Astrophysique, CP-226, Universit\'e Libre de Bruxelles, 
1050 Brussels, Belgium}
\author{N. G\"urlebeck}
\affiliation{Center of Applied Space Technology and Microgravity (ZARM), University of Bremen, 
Am Fallturm, 28359 Bremen, Germany}
\author{A. F. Fantina}
\affiliation{Institut d'Astronomie et d'Astrophysique, CP-226, Universit\'e Libre de Bruxelles, 
1050 Brussels, Belgium}
\affiliation{Grand Acc\'el\'erateur National d'Ions Lourds (GANIL), CEA/DRF - CNRS/IN2P3, Bvd Henri Becquerel, F-14076 Caen, France}
\author{J.~M.~Pearson}
\affiliation{D\'ept. de Physique, Universit\'e de Montr\'eal, Montr\'eal 
(Qu\'ebec), H3C 3J7 Canada}
\author{C. Ducoin}
\affiliation{Institut de Physique Nucl\'eaire de Lyon, Domaine scientifique de la Doua, B\^at. Paul Dirac, 
4 Rue Enrico Fermi, 69622 Villeurbanne Cedex, France.}

\begin{abstract} 
Giant pulsar frequency glitches as detected in the emblematic Vela pulsar have long been
thought to be the manifestation of a neutron superfluid permeating the inner crust of a neutron
star. However, this superfluid has been recently found to be entrained by the crust, and as a consequence 
it does not carry enough angular momentum to explain giant glitches. The extent to which 
pulsar-timing observations can be reconciled with the standard vortex-mediated glitch theory is studied considering the current uncertainties on 
dense-matter properties. To this end, the crustal moment of inertia of glitching pulsars is calculated employing 
a series of different unified dense-matter equations of state. 
\end{abstract}


\maketitle

\section{Introduction}
\label{intro}

Although most pulsars are spinning with an extremely stable frequency, some of them undergo abrupt changes. So far, 
472 such glitches have been detected in 165 pulsars~\cite{jod12}. One of the most emblematic glitching pulsars is Vela 
(PSR~B0833$-$45). Since its discovery in 1969, this pulsar has exhibited regular glitches with giant jumps of the angular 
velocity of order $\Delta\Omega\slash \Omega\sim 10^{-6}$. These spin-ups have been generally attributed to the sudden unpinning of 
neutron superfluid vortices in the inner crust of the neutron star~\cite{and75,alp85} (see, e.g. Section 12.4 of 
Ref.~\cite{lrr} for a review). Although different kinds of neutron superfluids are thought to exist in neutron-star cores, 
they were found to be strongly coupled to the crust due to mutual neutron-proton entrainment effects~\cite{alp84}. Neutron 
superfluidity, which was actually predicted before the discovery of pulsars~\cite{mig59}, has found additional support from 
observations of the initial cooling in transiently accreting neutron stars~\cite{shternin07,brown09}, from the rapid cooling 
of the neutron star in Cassiopeia A~\cite{pag11,sht11}, and from measurements of pulsar-braking indices~\cite{alp06,ho12} 
(for a recent review of neutron-star superfluidity, see, e.g. Ref.~\cite{page14}). 

The vortex-mediated glitch scenario has been recently challenged~\cite{cc06,and12,cha13,hoo15,new15,li15,li16} by the realization that 
despite the absence of viscous drag the neutron superfluid is still strongly coupled to the neutron-star crust due to Bragg 
scattering~\cite{cha05,cch05,cha06,cha12}. On the other hand, it has been argued that the crustal superfluid may carry enough 
angular momentum to explain Vela glitches invoking the lack of knowledge of the dense-matter equation of state (EoS)~\cite{piek14,ste15}. 
However, the EoSs employed in these analyses are not thermodynamically consistent, and more importantly, they are found to be incompatible with 
existing constraints from laboratory experiments and astrophysical observations. In particular, the authors of Ref.~\cite{piek14} determined 
the EoS of neutron stars using a relativistic mean-field model for the core, and the model of Baym, Pethick and Sutherland (BPS) for the outer crust~\cite{bps71}. 
The EoS of 
the inner crust was obtained from a polytropic interpolation. By fine-tuning the parameters of the relativistic mean-field Lagrangian, they showed that the 
thickness of the neutron-star crust could be adjusted so as to be compatible with observations of Vela pulsar glitches. Two parametrizations consistent with 
the existence of massive neutron stars such as PSR~J0348$+$0432~\cite{ant13} were thus found: NL3max and TFcmax. On the other hand, as pointed out by the authors 
of Ref.~\cite{piek14}, these parametrizations predict neutron-star radii ($R \gtrsim 14$~km for neutron stars with a mass $M=1.4 M_\odot$ where $M_\odot$ is the 
mass of the Sun) that are significantly larger than those inferred in Refs.~\cite{ste13,gui14,ozel15} from the analysis of X-ray bursters and quiescent low mass 
X-ray binaries. Although large neutron-star radii are not totally excluded by astrophysical observations (see, e.g. Ref.~\cite{sul11}; but see also the recent 
discussion in Ref.~\cite{nat15}), they are incompatible with experimental data and many-body calculations~\cite{lat13,heb13}. In addition, the symmetric nuclear-matter 
incompressibility coefficient at saturation predicted by NL3max, $K_v=271.54$~MeV, lies outside the range of values inferred from isoscalar giant monopole resonances 
in finite nuclei (see, e.g. Ref.~\cite{todd05}). As a matter of fact, the analysis of the matter flow and kaon production in heavy-ion collision 
experiments~\cite{dan02,lynch09} showed that the symmetric nuclear matter EoS predicted by NL3 (which is identical to that obtained with NL3max 
for the reasons discussed in Ref.~\cite{piek14}) is too stiff at any baryon number densities up to $4.5 n_0$, where $n_0\simeq 0.16$~fm$^{-3}$ is the 
density of symmetric nuclear matter at saturation. 
As for the parametrization TFcmax, the predicted values for the symmetry energy coefficient $J=38.3$~MeV and its slope $L=74.0$~MeV lie 
outside the range of values obtained from various experimental and theoretical constraints~\cite{tsa12,lat14}. Since the crust-core boundary is generally correlated 
with $L$~\cite{duc11}, the TFcmax parametrization may thus have lead to an overestimate of the crust inertia.  
In Ref.~\cite{ste15}, the uncertainties in the neutron-star properties were estimated from Monte Carlo simulations considering different 
parametrizations of the EoS. In particular, either a piecewise polytropic EoS or a four line segment equation 
of state was employed for the inner core. The EoS of the outer core was taken from quantum Monte Carlo calculations or chiral 
effective field theory calculations. Finally, a compressible liquid drop model was applied to determine the EoS of the crust. 
According to this study, the standard scenario of Vela pulsar glitches still remain viable considering the uncertainties in the dense matter 
EoS. On the other hand, this study also shows that the corresponding EoS are incompatible with existing neutron-star 
mass and radius measurements from X-ray bursters and quiescent low mass X-ray binaries, as also found in Ref.~\cite{piek14}. 

In this paper, the extent to which giant pulsar glitches can be explained by the superfluid in neutron-star crusts is more closely examined given the current 
uncertainties on dense-matter properties. To this end, we have employed a set of different unified EoSs. Treating all regions of a 
neutron star within the same nuclear model ensures that the resulting EoSs are thermodynamically consistent. This avoids the 
occurrence of spurious instabilities in neutron-star dynamical simulations. Moreover, this guarantees an accurate description 
of the transitions between the outer and inner parts of the crust, as well as the crust-core boundary. This is of utmost importance for the present study
since the ad hoc matching of differents EoSs for the crust and the core has been shown to induce large errors on the crust thickness~\cite{for2016}. 
The EoSs considered here are based on the nuclear energy density functional (EDF) theory 
using the accurately calibrated Brussels-Montreal EDFs BSk14~\cite{sg07}, BSk20-21~\cite{gcp10} and BSk22-26~\cite{gcp13}. 
For comparison, we have also adopted two other unified EoSs: SLy~\cite{dou01} and BCPM~\cite{bcpm}. The constraint on the global structure of a neutron star inferred 
from giant pulsar glitches will be briefly reviewed in Section~\ref{sec:constraint}. After presenting our models of rotating neutron stars in Section~\ref{sec:ns-model}, 
and the unified EoSs in Section~\ref{sec:eos}, results will be discussed in Section~\ref{sec:results}.

\section{Glitch constraint on the inertia of the crustal superfluid}
\label{sec:constraint}

Giant pulsar glitches as observed in Vela are generally thought to arise from sudden transfers of angular momentum between the neutron superfluid permeating 
the neutron-star crust and the rest of the star. Taking into account entrainment effects arising from Bragg scattering of unbound neutrons by neutron-proton 
clusters, the angular momentum $\mathcal{J}_{\rm s}$ of the superfluid depends on both the angular velocity $\Omega_{\rm s}$ of the superfluid and on the 
observed angular velocity $\Omega$ of the rest of the star~\cite{cc06}: 
\begin{equation}
\label{1}
\mathcal{J}_{\rm s}=I_{\rm ss} \Omega_{\rm s} + I_{\rm sc} \Omega\, ,
\end{equation}
where $I_{\rm ss}$ and $I_{\rm sc}$ are partial moments of inertia. Likewise, the angular momentum of the rest of the star can be expressed as~\cite{cc06} 
\begin{equation}\label{2}
\mathcal{J}_{\rm c}=I_{\rm sc} \Omega_{\rm s} + I_{\rm cc} \Omega \, . 
\end{equation}
The total angular momentum of the star can thus be written as 
\begin{equation}\label{3}
\mathcal{J}=\mathcal{J}_{\rm s}+\mathcal{J}_{\rm c}=I_{\rm s} \Omega_{\rm s} + I_{\rm c} \Omega \, ,
\end{equation}
where $I_{\rm s}=I_{\rm ss}+I_{\rm sc}$ and $I_{\rm c}=I_{\rm sc}+I_{\rm cc}$ are the moments of inertia of the superfluid and of the rest 
of the star respectively. 

Regardless of the actual glitch triggering mechanism, this two-component model leads to the following constraint~\cite{cc06}
\begin{equation}
\label{4}
\frac{(I_{\rm s})^2}{I I_{\rm ss}}\geq \mathcal{G} \, ,
\end{equation}
where $I=I_{\rm s}+I_{\rm c}$ is the total moment of inertia of the star. 
The coefficient $\mathcal{G}$ can be obtained from pulsar-timing data and is defined as
\begin{equation}
\label{5}
\mathcal{G}\equiv 2 \tau_c A_g \, ,
\end{equation}
where $\tau_c=\Omega/(2|\dot\Omega|)$ is the pulsar characteristic age 
and the glitch activity parameter $A_g$ is given by the sum over glitches occurring during a time $t$
\begin{equation}
\label{6}
A_g=\frac{1}{t}\sum_i\frac{\Delta\Omega_i}{\Omega}\, .
\end{equation}
Equation~(\ref{4}) can be alternatively expressed as~\cite{cc06} 
\begin{equation}
\label{7}
\frac{I_{\rm s}}{I}\geq \mathcal{G} \frac{\bar m_n^\star}{m_n} \, ,
\end{equation}
where $\bar m_n^\star$ is a suitably weighted mean of the local effective neutron mass $m_n^\star$ introduced in Refs.~\cite{cha05,cch05}. 
Comparing Eqs.~(\ref{4}) and (\ref{7}) thus yields 
\begin{equation}
 \frac{\bar m_n^\star}{m_n} \equiv \frac{I_{\rm ss}}{I_{\rm s}} \, .
\end{equation}
This mean effective neutron mass depends mainly on the properties of the neutron-star crust, as shown below. 
Values of the coefficient $\mathcal{G}$ for different glitching pulsars can be found, e.g., in Ref.~\cite{ho15}. For Vela, this coefficient is 
$\mathcal{G}=1.62\pm0.03\%$, thus leading to the following constraint:  
\begin{equation}
\label{eq:glitch-constraint-Vela}
\frac{I_{\rm crust}}{I}\biggr\vert_{\rm Vela} \geq \frac{I_{\rm s}}{I}\biggr\vert_{\rm Vela}\geq \frac{\bar m_n^\star}{m_n} (1.62\pm0.03)\times 10^{-2} \, ,
\end{equation}
where $I_{\rm crust}$ denotes the moment of inertia of the crust. 
Other glitching pulsars such as PSR~B1800$-$21 and PSR~B1930$+$22 exhibit higher values for the coefficient $\mathcal{G}$~\cite{and12,ho15}. On the other hand, the estimated 
errors are also larger. For this reason, these pulsars will not be considered in this study.

The fractional moment of inertia of the crust and that of the crustal superfluid depend on the global structure of the neutron star, which we have computed 
by solving numerically Einstein's equations of general relativity, as described in the following Section.

\section{Neutron-Star Model}
\label{sec:ns-model}

The equations describing the global structure of a rotating neutron star can be obtained from Einstein's equation and the conservation of the stress energy-density 
tensor $T^{ab}$:
\be
G_{ab} = 8\pi T_{ab},\ \nabla_a T^{ab} = 0\, ,
\ee
where $a$ and $b$ denote space-time indices, and $G_{ab}$ is the Einstein's tensor. 
Note that throughout this paper we adopt units such that $G=c=1$, where $G$ is the gravitational constant and $c$ is the speed of light. 
In the following, we shall consider stationary and axially symmetric rigidly rotating stellar configurations. 

As in previous studies~\cite{and12,cha13,hoo15,new15,li15,li16,piek14,ste15}, we shall compute the moments of inertia assuming that the superfluid is corotating 
with the rest of star, i.e. neglecting the small difference $|\Omega_{\rm s}-\Omega|\ll \Omega$. 
The stress energy-density tensor will thus be taken to be that of a perfect fluid:
\be
T_{ab} = (\rho + P) u_a u_b + P g_{ab},
\ee
where $\rho$ is the mass-energy density, $P$ the pressure, $g$ the metric and $u$ the $4$-velocity of the fluid normalized as $u_a u^a =-1$. 
We shall also consider the limit of slow rotation. The ensuing form of Einstein's equations, and the expressions for the partial moments of inertia are given in Section~\ref{sec:slow-rot}. The validity of this approximation will be assessed by extending our neutron-star models to rapid rotations. The formalism is briefly 
discussed in Section~\ref{sec:rr}.

\subsection{Slow-rotation approximation}
\label{sec:slow-rot}

The metric describing a slowly rotating neutron star is approximately given by 
(see, e.g. Ref.~\cite{fried13}) 
\begin{align}
ds^2&=-b(r) dt^2  +\left(1-\frac{2 m(r)}{r}\right)^{-1} dr^2 \nonumber\\
   &+r^2  \Biggl[d\theta^2+\sin^2\theta  \biggl(d\varphi- \epsilon  \left(\Omega-\omega_1(r)\right)dt\biggr)^2\Biggr],
\end{align}
where $b(r)$, $m(r)$, and $\omega_1(r)$ are functions of the radial coordinate $r$, $\theta$ and $\varphi$  are the polar and azimuthal angles respectively, 
$\Omega$ is the neutron-star angular velocity as measured by a distant observer, and $\epsilon$ is a small dimensionless parameter. 

We expand Einstein's equations in powers of $\epsilon$, following the procedure introduced by Hartle~\cite{hartle}~:
\begin{align}\label{eq:hartle}
&m' = 4\pi r^2 \rho,\nonumber\\
&\frac{b'}{b} = 2\left( \frac{m + 4\pi r^3 P}{r(r-2m)} \right),\nonumber\\
&P' = \frac{m + 4\pi r^3 P}{r(r-2m)}(\rho + P),\nonumber\\
&\omega_1'' = \frac{16 \pi r^2 (\rho + P)\omega_1 + (4(2m-r) + 4\pi r^3 (\rho + P) )\omega_1'}{r (r-2m)}\, ,
\end{align}
where a prime denotes the derivative with respect to $r$. 
This system of equations is solved with the following boundary conditions
\be\label{eq:bc}
m(0)=0,\ P(0)=P_0,\ b(0)=b_0,\ \omega_1(0) = 1,\ \omega_1'(0)=0,
\ee
where $P_0$ is the central pressure of the star and the conditions on $\omega_1$ follow from elementary flatness and the symmetry. 
Note that the (linear) equation of $\omega_1$ is invariant under rescalings $\omega_1 \rightarrow \lambda \omega_1$, allowing us to set $\omega_1(0)=1$ 
without loss of generality. The value of $b_0$ is obtained by the requirement that the function $b$ matches continuously the exterior spacetime described by 
the Schwarzschild solution
\be
b(r)  = 1-\frac{2M}{r}, \ m(r) = M,
\ee
where $M$ is the gravitational mass of the star, 
defined by the function $m(r)$ evaluated at the radius $R$ for which the pressure vanishes:
\be
P(R)=0,\ M = m(R)\, .
\ee
In order to complete the model, an EoS must be specified. The EoSs adopted in this work will be discussed 
in Section~\ref{sec:eos}.

The total momentum of inertia $I$ of the slowly rotating star can be obtained by matching the function $\omega_1$ to the exterior spacetime, 
described by
\be
\omega_1  = \Omega\left(1-\frac{2I}{r^3}\right).
\ee
Note that the value of the inertia is independent of the boundary condition on $\omega_1$, see the remark below \eqref{eq:bc}. For the same argument, it is independent 
of the value of $\Omega$.

It can be shown using Einstein's field equations that the total moment of inertia can be equivalently expressed as (see, e.g., Ref.~\cite{lat01})
\be 
I=I(R)\, , 
\ee 
with 
\begin{eqnarray}
I(r) =&& \frac{8\pi}{3}\int_0^r x^4  n(x) \mu(x)\frac{\omega_1(x)}{\Omega}\nonumber \\ 
 && \times \sqrt{b(x)\left(1-\frac{2m(x)}{x}\right)} dx\, ,
\end{eqnarray}
where $n$ is the average baryon number density, and $\mu$ the corresponding baryon chemical potential. 
The fractional moment of inertia of the crust is given by 
\be\label{eq:Icrust}
\frac{I_{\rm crust}}{I}  = 1-\frac{I(r_{\rm core})}{I}\, , 
\ee
and $r_{\rm core}$ denotes the radial coordinate at the crust-core boundary where the pressure is $P_{\rm core}$ (as determined by dense-matter models)~:
\be
P(r_{\rm core}) = P_{\rm core} \, .
\ee
The partial and total moments of inertia of the crustal superfluid are given by~\cite{cc06} 
\begin{eqnarray}\label{eq:Is}
I_{\rm s} =&& \frac{8\pi}{3}\int_{r_{\rm core}}^{r_{\rm drip}} x^4  n_n^{\rm f}(x) m_n \frac{\omega_1(x)}{\Omega}\nonumber \\ 
 && \times \sqrt{b(x)\left(1-\frac{2m(x)}{x}\right)} dx\, ,
\end{eqnarray}
\begin{eqnarray}\label{eq:Iss}
I_{\rm ss} =&& \frac{8\pi}{3}\int_{r_{\rm core}}^{r_{\rm drip}} x^4  n_n^{\rm f}(x) m_n^\star(x) \frac{\omega_1(x)}{\Omega}\nonumber \\ 
 && \times \sqrt{b(x)\left(1-\frac{2m(x)}{x}\right)} dx\, ,
\end{eqnarray}
where $r_{\rm drip}$ is the radial coordinate at the neutron-drip transition, $n_n^{\rm f}$ the number density of 
free neutrons, and $m_n^\star$ their effective mass, as defined in Ref.~\cite{cha12}. 
In the slow-rotation approximation, $I_{\rm crust}$, $I_{\rm ss}$ and $I_{\rm s}$ are 
independent of the angular frequency. This is, however, not anymore true in the case of rapid rotation and a quadratic increase is 
expected for the next order in the expansion parameter $\epsilon$.

 \subsection{Rapid rotations}
 \label{sec:rr}

If the neutron star is rotating rapidly enough, higher-order corrections to the slow-rotation approximation could in principle increase the fractional moment of inertia 
of the crust. To assess the importance of these corrections, we performed a numerical analysis 
using a version of the \emph{rns} code~\cite{Ste,Ste95}, which was modified by J.\ Steinhoff and applied in Ref.~\cite{Cha2014}. 
In the \emph{rns} code, the spacetime metric is expressed in the form
\begin{align}
\begin{split}
ds^2=-&\mathrm{e}^{\gamma+\beta} dt^2 + \mathrm{e}^{2\alpha}(dr^2+r^2 d\theta^2)+\\
&\mathrm{e}^{\gamma-\beta}r^2\sin^2\theta\left(d\varphi-\omega dt\right)^2,
\end{split}
\end{align}
where $\gamma,~\beta,~\alpha$ and $\omega$ are functions solely of $\theta$ and $r$. 

The moment of inertia of the star is given by 
\begin{equation}\label{eq:MoI}
I=\frac{\mathcal{J}}{\Omega}\, .
\end{equation}
The angular momentum $\mathcal{J}$ can be determined either from the asymptotic behavior of the metric function $\omega$ or from the Komar integral~\cite{Kom59}. 
The approach based on the Komar integral can be seen as a source integral formalism for the first multipole moments, as described in Ref.~\cite{Gur14}. It can be easily employed to assign to any region of the neutron star the associated angular momentum. For example, the angular momentum contained in the crust is determined by~\cite{Bar73}

\begin{equation}\label{eq:Jdef}
\mathcal{J}_{\rm crust}=2\pi\int\limits_{V} n\mu\, {\rm e}^{\gamma-\beta+2\alpha} \frac{v}{1-v^2} r^3\sin^2\theta  d\theta dr.
\end{equation}
$V$ is the coordinate volume of the crust and $v$ is the the proper velocity of a fluid element with respect to a local zero angular momentum observer, given by 
\begin{align}
v={\rm e}^{-\beta} r \sin\theta (\Omega-\omega).
\end{align}

Using \eqref{eq:MoI}, the fractional moment of inertia of the crust is thus finally given by
\begin{align}
\frac{I_{\rm crust}}{I}=\frac{\mathcal{J}_{crust}}{\mathcal{J}}\, .
\end{align}

\section{Unified equations of state of dense matter} 
\label{sec:eos}

The interior of a neutron star can be decomposed into several qualitatively distinct regions, which can be classified as
follows with increasing depth: 
i) A very thin atmospheric plasma layer of light elements (mainly hydrogen and helium though heavier elements 
like carbon may also be present) possibly surrounds a Coulomb liquid of electrons and ions,  
ii) the \emph{outer crust} consists of free electrons and pressure-ionized atoms arranged in a body-centered 
cubic lattice (see, e.g., Refs.~\cite{pea11,roca2008,hemp13}), iii) the appearance of free neutrons at densities above 
$\rho_{\rm drip} \approx 4 \times 10^{11}$~g~cm$^{-3}$ (see, e.~g. Ref.~\cite{cfzh15} for a detailed discussion) marks 
the transition to the \emph{inner crust}, which extends up to about $\rho_{\rm core}\approx 10^{14}$~g~cm$^{-3}$ (see, e.g. Ref.~\cite{lrr}), 
iv) the \emph{outer core} made of nucleons and leptons up to densities $2-3 \rho_0$, and v) the inner core, 
whose composition remains highly uncertain. For simplicity, we assume that the inner core consists of nucleons 
and leptons only. As we shall see in Section~\ref{sec:results}, the density at the center of the Vela pulsar, as 
inferred from the glitch data, actually lies below $\sim 2-3 \rho_0$. 

In the following, we shall adopt the cold catalyzed matter hypothesis according to which matter in neutron-star interiors are in full 
thermodynamic equilibrium at zero temperature. In this case, the surface layers are solid and consist of $^{56}$Fe up to densities of 
about $8\times 10^6$ g~cm$^{-3}$. For this region, we employ the EoSs from Table 5 of Ref.~\cite{lai91}: the one referred to as ``QEOS'' 
for densities below $1.4\times 10^3$~g~cm$^{-3}$ (this EoS was found to be in good agreement with experiments~\cite{bat02}), and the 
one referred to as ``TFD'' up to about $4.3\times 10^5$~g~cm$^{-3}$. For the denser regions of the star, the composition and the EoS depends 
on the nuclear model. Although the different states of dense matter encountered in this interval require different types of treatments 
(as described below, see also Ref.~\cite{fant15}), these calculations are performed using the \emph{same} EDF. 
The transitions are thus described self-consistently.
The composition and the EoS of the outer crust at densities above $4.3\times 10^5$~g~cm$^{-3}$ were calculated in the framework of the BPS model~\cite{bps71}, 
making use of the latest experimental atomic mass data complemented with the Hartree-Fock-Bogoliubov (HFB) predictions for masses that have not 
yet been measured (see Ref.~\cite{pea11} for details). For the inner crust, the 4th order extended Thomas-Fermi method was employed. 
Proton shell corrections were included via the Strutinsky integral theorem (neutron shell effects are negligibly small except possibly 
near the neutron-drip point). This so-called ETFSI method (extended Thomas-Fermi+Strutinsky integral) is a computationally 
high-speed approximation to the fully self-consistent Hartree-Fock equations~\cite{pea12}. Proton pairing was included as in 
Ref.~\cite{pea15}. In order to further optimize the computations, the nucleon density distributions were parameterized, and we adopted 
the spherical Wigner-Seitz approximation to calculate the Coulomb energy, since nuclear clusters are essentially spherical, except possibly 
near the crust-core interface where so-called nuclear ``pastas''  might exist~\cite{lrr}.
Even though two different codes were used to calculate the EoS in the outer crust and in the inner crust, the pressure (energy per nucleon) 
at the boundary was found to differ by less than 3\% (5\%) thus ensuring a thermodynamically consistent description of the EoS~\cite{pea12}.  
Deeper in the star, the core is described by the EoS of homogeneous beta-equilibrated matter made of nucleons and leptons (including muons 
at high density). 
The crust-core transition can be determined by the instability of such homogeneous matter against density fluctuations. The transition density 
and pressure were calculated by the method described in Ref.\cite{duc07}, which was shown to be extremely accurate~\cite{pea12}.

In this way, unified EoSs for neutron stars were calculated using the latest series of Brussels-Montreal (BSk) EDFs. These 
EDFs were constructed from generalized Skyrme effective interactions (see, e.g. Ref.~\cite{cha15} for a brief overview). The parameters 
of these interactions were determined primarily by fitting essentially all measured atomic masses, which were calculated 
using the HFB method. In particular, the root-mean square deviation between the measured atomic 
masses with neutron number $N \geq 8$ and proton $Z \geq 8$ from the 2012 Atomic Mass Evaluation~\cite{ame12} and the 
theoretical HFB masses amounts to about $0.5$~MeV for the latest EDFs. A set of different EDFs were constructed 
by imposing different values for the symmetry energy $J$ at saturation, and by fitting different realistic neutron-matter 
EoSs spanning different degrees of stiffness corresponding to current predictions of various microscopic calculations. 
Moreover, these EDFs were fitted to realistic $^1S_0$ pairing gaps in neutron matter and in symmetric nuclear matter. A number 
of additional constraints were imposed to the Brussels-Montreal EDFs: i) the incompressibility $K_v$ of symmetric nuclear matter 
at saturation was required to fall in the empirical range $240 \pm 10$~MeV \cite{colo04}, ii) the isoscalar effective mass in symmetric 
nuclear matter at saturation was fixed to the realistic value of $0.8$ (see Ref.~\cite{gcp10} for a summary of the experimental and 
theoretical evidence), iii) the parameters of the interactions were adjusted so as to limit the occurence of spurious spin and spin-isospin 
instabilities, and in particular to prevent a ferromagnetic collapse of neutron stars. Although the Brussels-Montreal EDFs were not directly 
fitted to realistic EoS of symmetric nuclear matter, they are compatible with the constraints inferred from the analysis of heavy-ion 
collision experiments~\cite{dan02,lynch09}. Besides, the predicted values for the symmetry energy $J$ and its slope $L$ at saturation 
are consistent with those inferred from from various experimental and theoretical constraints~\cite{tsa12,lat14}. Finally, the isovector 
effective mass was found to be smaller than the isoscalar effective mass at the saturation density. This result is consistent with measurements 
of isovector giant resonances~\cite{les06}, and microscopic calculations~\cite{van05,zuo06}. 
These features make the Brussels-Montreal EDFs well-suited for a unified treatment of neutron stars. For the BSk20-26 versions~\cite{gcp10,gcp13}
we adopt in this work, unified EoSs have been calculated~\cite{pea11,pea12,pea15,pcfg14} and tested against astrophysical observations~\cite{fant13,fant15}.
Analytical representations are available for the unified EoSs based on the EDFs BSk20-21~\cite{pot2013}.

For comparison, we have also considered two other unified EoSs based on the EDF theory: SLy~\cite{dou01} and BCPM~\cite{bcpm}. 
The SLy EoS does not actually provide a fully consistent description of neutron stars: indeed, the EoS of the outer crust was not calculated using the same 
EDF, but was taken from Ref.~\cite{hp94}. The reason may lie in the fact that HFB calculations using the SLy4~\cite{chab98,chab98e} EDF that underlies the SLy EoS, 
yields a rather poor fit to experimental nuclear masses, with a root-mean-square deviation (considering only even-even nuclei) of about 5.1 MeV~\cite{dsn04}. 
The EoS of the inner crust was calculated using the compressible liquid drop model. In this 
approximation, neutron-proton clusters are assumed to have sharp surfaces. Moreover, nucleons in clusters and unbound nucleons are treated differently and 
shell effects are ignored. The EDF underlying the BCPM EoS was constructed by performing an educated polynomial fit of microscopic calculations in infinite 
homogeneous nuclear matter using realistic nucleon-nucleon potentials. The EDF was supplemented with additional phenomenological surface and spin-orbit 
terms that were fitted to properties of finite nuclei. The BPS model was employed to compute the EoS of the outer crust 
making use of experimental masses as well as theoretical masses obtained from HFB calculations. The EoS of the inner crust was determined using the simple Thomas-Fermi  
method, without higher-order corrections; neither pairing nor shell corrections were included. On the other hand, contrary to the Brussels-Montreal EoSs, nucleon density distributions were not parametrized, but were calculated self-consistently allowing for nuclear pastas.

The evaluation of the partial moment of inertia $I_{\rm ss}$ requires the knowledge of the local neutron effective mass $m_n^\star$ in all regions of the inner crust
of a neutron star. However, $m_n^\star$ has not been calculated for the unified EoSs considered here. As a matter of fact, systematic calculations of $m_n^\star$ are 
computationally extremely costly, and for this reason were only performed in Ref.~\cite{cha12} based on the crustal composition taken from Ref.~\cite{onsi08} using the 
Brussels-Montreal EDF BSk14~\cite{sg07}. Therefore, we have computed the ratios $I_{\rm ss}/I_{\rm crust}$ and $I_{\rm s}/I_{\rm crust}$ using the EoS based on the EDF BSk14, and kept the resulting values for all the other unified EoSs. To this end, we have constructed a unified EoS with BSk14. The EoS of the outer crust was determined as in Ref.~\cite{pea11},
and that of the core was calculated considering homogeneous matter in beta equilibrium. Like all the other BSk EDFs, BSk14 was fitted to all experimental atomic 
masses with $Z,N\geq 8$ (masses from the 2003 Atomic Mass Evaluation~\cite{ame03} were fitted with a root-mean-square deviation of 0.73 MeV), but was constrained to reproduce a softer neutron-matter EoS than those considered in the more recent series of Brussels-Montreal EDFs.

The crust-core transition density and pressure are indicated in Table~\ref{tab1} for all the EoSs considered 
in this study. 

\begin{table}
\caption{Average baryon number density $n_{\rm core}$ and pressure $P_{\rm core}$ at the crust-core interface, 
as predicted by different nuclear energy density functionals. See text for details.}
\label{tab1}
\begin{tabular}{ccc}
\hline 
 & $n_{\rm core}$ (fm$^{-3}$)& $P_{\rm core}$ (MeV fm$^{-3}$)\\
\hline 
BSk14 & 0.0810 & 0.381 \\
BSk20 & 0.0854& 0.365 \\
BSk21 & 0.0809& 0.269\\
BSk22 & 0.0716 & 0.291 \\
BSk24 & 0.0808 & 0.268 \\
BSk25 & 0.0856 & 0.211 \\
BSk26 & 0.0849 & 0.363 \\
SLy4 & 0.0798 & 0.361 \\
BCPM  & 0.0825 & 0.432 \\
\hline
\end{tabular}
\end{table}

\section{Results and discussion}
\label{sec:results}

\begin{figure}
 \centering
 \includegraphics[scale=.35]{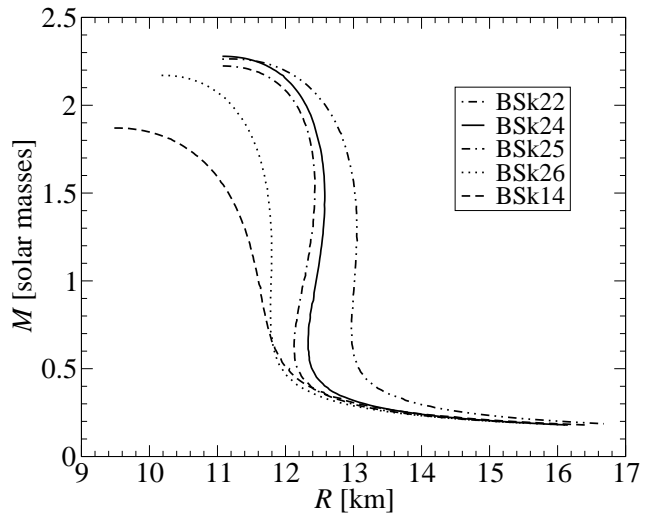}
 \caption{Masses and radii of nonrotating and nonaccreting neutron stars, as obtained for different unified Brussels-Montreal equations of state based on the following 
 energy-density functionals:  BSk14~\cite{onsi08,sg07}, BSk22, BSk24, BSk25 and BSk26~\cite{gcp13,pcfg14}.}
 \label{fig:MR_BSk}
\end{figure}

\begin{figure}
 \centering
 \includegraphics[scale=.35]{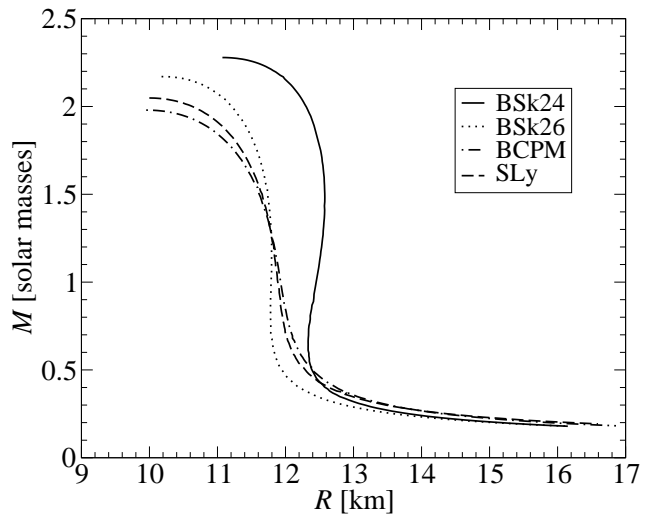}
 \caption{Same as Fig.~\ref{fig:MR_BSk} for other unified equations of state: SLy~\cite{dou01}, and BCPM~\cite{bcpm}.}
 \label{fig:MR}
\end{figure}

\begin{figure}
 \centering
 \includegraphics[scale=.35]{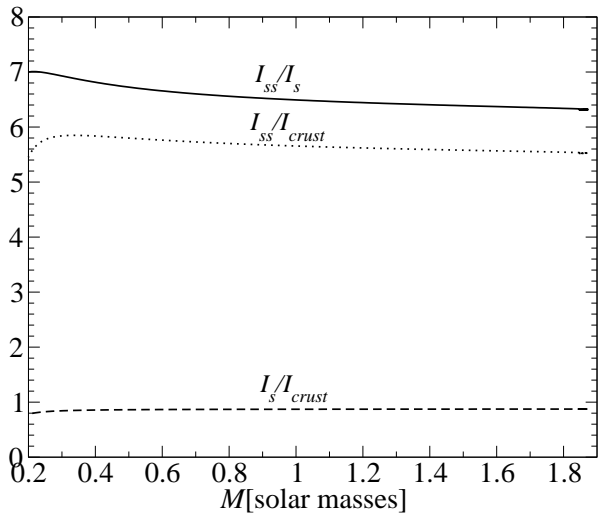}
 \caption{Fractional moments of inertia of a neutron star as a function of the neutron-star mass, as obtained using the unified equation of state based on 
 the energy-density functional BSk14~\cite{sg07}. See text for details.}
 \label{fig:Iratios}
\end{figure}

For each EoS described in Section~\ref{sec:eos}, we have computed the set of stationarily rotating neutron-star configurations by solving 
numerically Eqs.~(\ref{eq:hartle}). In the slow-rotation approximation, the structure of the star is the same as that of a static star. The corresponding 
masses and radii are shown in Figs.~\ref{fig:MR_BSk} and \ref{fig:MR}. The mass-radius curves obtained with the EoSs based on BSk20 and BSk21 are undistinguishable from 
those obtained with the EoSs based on BSk26 and BSk24 respectively. The results agree with those previously obtained in Refs.~\cite{fant13,fant15}. The EoS based on BSk14 predicts a maximum mass significantly lower than that obtained with the other Brussels-Montreal EDFs, and below the measured mass $2.01\pm 0.04~M_\odot$ 
of PSR~J0348+0432~\cite{ant13}. This stems from the softer neutron-matter EoS to which this EDF was fitted, namely that calculated by 
Friedman and Pandharipande using realistic two- and three-body forces~\cite{fp81}. Nevertheless, this EoS remains compatible with more recent 
ab initio calculations at densities of relevance for the crust and the outer core of neutron stars~\cite{tew13}. For this reason, we believe that 
BSk14 is still suitable for calculating the fractional moments of inertia $I_{\rm ss}/I_{\rm crust}$ and $I_{\rm s}/I_{\rm crust}$. Indeed, these 
ratios depend mainly on the properties of the neutron-star crust. They were previously estimated within the thin-crust approximation as 
$I_{\rm ss}\approx 4.6 I_{\rm crust}$ and $I_{\rm s}\approx 0.89 I_{\rm crust}$~\cite{cha13}. The associated mean effective neutron mass 
is about $\bar m_n^\star \approx 5.1 m_n$ (values $\sim 14-16\%$ lower were found in Ref.~\cite{and12}, but calculations were not performed using 
the EoS based on BSk14). We have recalculated these quantities using 
Eqs.~(\ref{eq:Icrust}), (\ref{eq:Iss}) and (\ref{eq:Is}) by solving numerically Einstein's equations, as described in Section~\ref{sec:ns-model}. 
As shown in Fig.~\ref{fig:Iratios}, $I_{\rm ss}/I_{\rm crust}$ and $I_{\rm s}/I_{\rm crust}$ are almost independent of the global neutron-star 
structure. While the ratio $I_{\rm s}/I_{\rm crust}$ we find is in very good agreement with the value obtained within the thin-crust approximation, 
the ratio $I_{\rm ss}/I_{\rm crust}$ is about $20\%$ higher thus leading to a more stringent constraint. The origin of this discrepancy most presumably 
lies in the large variations of $m_n^\star$ in different crustal layers. In all these calculation, vortex pinning is supposed to be effective in all 
regions of the inner crust (in reality, the inertia of the neutron superfluid could be lower).

\begin{figure}
 \centering
 \includegraphics[scale=.35]{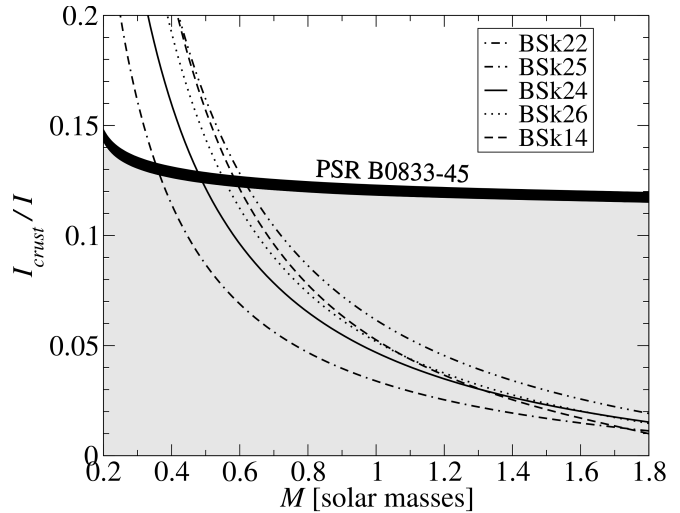}
 \caption{Fractional moment of inertia of the crust of a nonaccreting neutron star, as obtained for unified Brussels-Montreal equations of state based on the following 
 energy-density functionals: BSk14~\cite{sg07}, BSk22, BSk24, 
 BSk25 and BSk26~\cite{gcp13,pcfg14}. The light grey  region below the dark thick line is excluded by pulsar timing-data if giant glitches in Vela (PSR~B0833$-$45) originate from neutron-star crusts only. See text for details.}
 \label{fig:Icrust_BSk}
\end{figure}

\begin{figure}
 \centering
  \includegraphics[scale=.35]{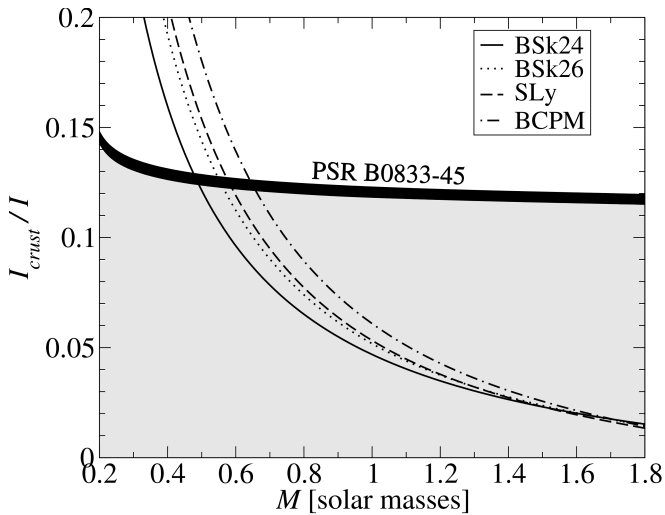}
 \caption{Same as Fig.~\ref{fig:Icrust_BSk} for other unified equations of state: SLy~\cite{dou01}, BCPM~\cite{bcpm}. The equations of state based on the energy-density functionals BSk24 and BSk26~\cite{gcp13,pcfg14} are plotted for comparison.}
 \label{fig:Icrust_other}
\end{figure}

We have calculated numerically the fractional moment of inertia of the crust of a neutron star with different masses using the unified EoSs presented in 
Section~\ref{sec:eos}. The results shown in Figs.~\ref{fig:Icrust_BSk} and \ref{fig:Icrust_other} can be easily 
analyzed after remarking that for a given mass $M$, $I_{\rm crust}/I$ is generally an increasing function of the radius $R$ and of the crust-core transition 
pressure $P_{\rm core}$ (see, e.g., Ref.~\cite{lat01}). In turn, the uncertainties in $P_{\rm core}$ arise from the lack of knowledge of the symmetry energy and 
its density dependence (see, e.g., Refs.\cite{horowitz2001,oyamatsu2007,roca2008,vidana2009,ducoin2011,grill2012,newton2013,sulaksono2014,provid2014,bao2014,
grill2014,seif2014,provid2014}). The dependence of $I_{\rm crust}/I$ on $R$ and $P_{\rm core}$ is best seen by comparing the results obtained for the EoSs based on 
the EDFs BSk22, BSk24, and BSk25. The EDFs underlying these unified EoSs were fitted to the same realistic neutron-matter EoS, but with different values for the symmetry energy 
coefficient, from $J=32$~MeV for BSk22 to $J=29$~MeV for BSk25, while $J=30$~MeV for BSk24. The highest values for $R$ and $P_{\rm core}$ are obtained for the EoS based on BSk22, the lowest for the EoS based on BSk25, the EoS based on BSk24 yielding intermediate values. As expected, the crustal moment of inertia follows the same hierarchy, with the largest values obtained with the EoS based on BSk22. The BSk26 EDF yields the same symmetry energy coefficient $J=30$~MeV as BSk24, but was fitted to a neutron-matter EoS that is softer at high densities. 
With a lower pressure to resist the gravitational pull, the unified EoS based on this EDF thus predicts smaller neutron stars. On the other hand, the EoS based on BSk26 predicts a higher crust-core transition pressure than the EoS based on BSk24. All in all, the crustal moment of inertia obtained with the EoS based on BSk26 is not much different from that obtained with the EoS based on BSk24, as shown in Fig.~\ref{fig:Icrust_BSk}. We have also found that the EoS based on BSk20 (BSk21) yields essentially the same crustal moment of inertia as the EoS based on BSk26 (BSk24). The corresponding curves are undistinguishable in Fig.~\ref{fig:Icrust_BSk}. The results obtained with the 
SLy and BCPM EoSs are close to those obtained with the Brussels-Montreal EoSs, as shown in Fig.~\ref{fig:Icrust_other}.

Assuming that only the neutron superfluid in the crust of a neutron star is responsible for the observed giant glitches leads to the constraint shown in 
Figs.~\ref{fig:Icrust_BSk} and \ref{fig:Icrust_other}. 
The inferred mass of the Vela pulsar is at most $M\simeq 0.66 M_\odot$. 
Such a low mass challenges the current scenarios of neutron-star formation (see, e.g., Section 3.3 in  Ref.~\cite{lat12} for a short review). Although neutron stars with a mass $M\sim M_\odot$ might be produced via fragmentation, they are not expected to be seen as ordinary radio pulsars (see, e.g. Ref.~\cite{pop07}). 
Besides, the comparison of neutron-star cooling simulations with the observational
estimates of the age and thermal luminosity of the Vela pulsar suggests that this neutron star is rather massive (see, e.g., Ref.~\cite{pot15}). In other words, the glitch 
puzzle still remains considering the current uncertainties in the dense-matter EoS. The reason lies in the fact that the average baryon number density inferred at the center 
of Vela is rather low, \emph{at most} $0.23-0.33$~fm$^{-3}$  depending on the EoS. At such densities, the neutron-star core is generally expected to contain nucleons and leptons only (see, e.g. Ref.~\cite{deb15} and references therein), and the dense-matter EoS is fairly well constrained by laboratory experiments, especially heavy-ion collisions~\cite{dan02,lynch09}. 


In this analysis, we considered that the fractional moments of inertia $I_{\rm s}/I_{\rm crust}$ and $I_{\rm ss}/I_{\rm crust}$ are the same for all EoSs, as in previous  
studies~\cite{and12,cha13,hoo15,new15,li15,li16,piek14,ste15}. Although the threshold density for the onset of the neutron-drip transition is model-dependent, the variations are quite small (see, e.g., Ref.~\cite{fant16}). The ensuying changes in $I_{\rm s}$ are expected to be negligible since $I_{\rm s}$ is mainly determined by the inertia of the denser regions of the inner crust. Ignoring the model dependence of $I_{\rm s}/I_{\rm crust}$ and $I_{\rm ss}/I_{\rm crust}$ seems also to be a reasonable approximation in view of the published results for the effective neutron mass in the inner crust of a neutron star~\cite{cha05,cha06,cha13}. 
For consistency and completeness, we have computed the global structure of rotating neutron stars using the unified EoS constructed from BSk14. 
As shown in Fig.~\ref{fig:Icrust_BSk}, BSk14 yields similar results for the ratio $I_{\rm crust}/I$ as those obtained with the more recent Brussels-Montreal EDF BSk26, 
as well as with the SLy EoS. 

Since the assumption of slow rotation may underestimate $I_{\rm crust}/I$, we have assessed the validity of this approximation 
for the EoS based on BSk24 using the \emph{rns} code. We have calculated different models of neutron stars varying the central density and the ratio of the polar to equatorial coordinate radius $r_e$. For neutron stars with a spin parameter $\mathcal{J}/M^2$ below $0.15$ we have reduced the polynomial order of the expansion in the \emph{rns} code to $3$, since higher orders contributed only numerical noise as in Ref.~\cite{Cha2014}. We chose a grid of $401\times 801$ in the compactified coordinate plane $[\cos\theta, r/(r+r_e)]$. A further increase of the grid size did not yield any improvements.
As a consistency check of the numerical method, we have calculated the total angular momentum via the two ways described in Section~\ref{sec:constraint}. The results have been found to deviate by less than $1\%$ for neutron-stars with a mass exceeding $0.5~M_\odot$, which is in accordance with previous comparative studies \cite{Ste95,Noz98}. Thus, an accuracy of approximately that order must be accepted for local quantities like $I_{crust}/I$, too. 

\begin{figure}
\centering
\includegraphics[scale=.35]{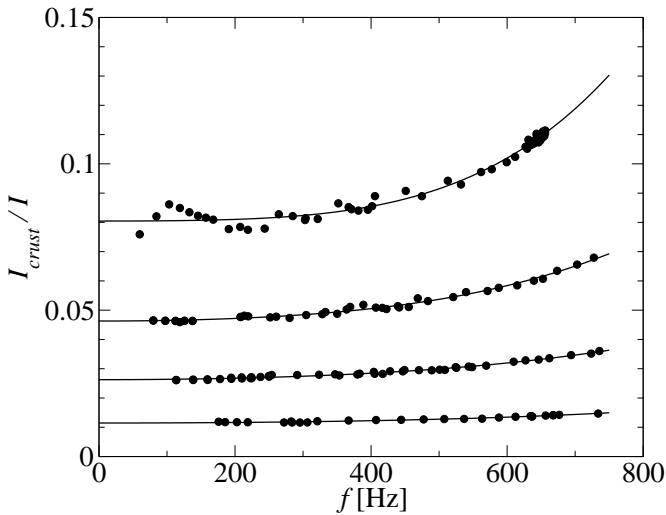}
\caption{Fractional moment of inertia of the crust of a nonaccreting neutron star as a function of the neutron star's rotation frequency $f=\Omega/(2\pi)$, as obtained using the unified equation of state based on the energy-density functional BSk24~\cite{gcp13}. Results are shown for four different neutron-star masses: $0.7$, $1$, $1.4$ and $2$ solar masses, respectively (from top to bottom). The points represent the actual numerical results and the curves are polynomial fits up to the fourth order.}
\label{fig:IcrustRatio_BSK24}
\end{figure}

In Fig.~\ref{fig:IcrustRatio_BSK24}, we exemplify the behavior of $I_{crust}/I$  for four different masses and varying frequency of the neutron star. Indeed, this ratio increases with the spin rate of the pulsar. Taking as an example the fastest spinning pulsar (PSR J1748-2446ad) known to date with a rotation frequency $f=\Omega/(2\pi)$ of about $716\,$Hz \cite{Hes06}, the relative difference of the slow-rotation approximation and the numerical results are between $23\%$ for a $2~M_\odot$ star and $36\%$ for a $0.7~M_\odot$ star, as shown in Fig.~\ref{fig:IcrustRatio_BSK24}. In contrast, the slow-rotation approximation is very accurate for stars spinning as slow as Vela ($f=11.195\,$Hz), for which the relative difference to the results obtained with the \emph{rns} code is far below the numerical noise limit of $1\%$. Nonetheless, the fits shown in Fig.~\ref{fig:IcrustRatio_BSK24} include already terms from an expansion in $f$ to the fourth order. In fact, the highest order is only relevant, i.e. above the 
numerical noise 
limit, for the $0.7~M_\odot$ stars ($6\%$ contribution). The fits reproduce also the results of the slow rotation approximation with an accuracy of $1-2\%$, which corroborates the expected accuracy regime.  

\section{Conclusions}

We have computed the global structure of rotating neutron stars, both in full general relativity and in the slow-rotation approximation. For this purpose, 
we have employed a series of unified EoSs, treating consistently all regions of a neutron star. The Brussels-Montreal unified EoSs presented in Ref.~\cite{pcfg14} 
are all based on the very accurately calibrated EDFs BSk22-26 spanning the current uncertainties in the nuclear symmetry energy as 
well as in the high-density 
stiffness of the neutron-matter EoS~\cite{gcp13}. For comparison, we have also considered the unified EoS based on BSk14~\cite{onsi08} (for which crustal entrainment 
was calculated in Ref.~\cite{cha12}), BSk20-21~\cite{pea11,pea12,pot2013}, SLy~\cite{dou01} and BCPM~\cite{bcpm}. In all cases, we find that the neutron superfluid permeating the inner crust of a neutron star does not carry enough angular momentum to explain the giant frequency glitches observed in the Vela pulsar. The glitch puzzle 
is not restricted to Vela, but concerns other glitching pulsars. On the other hand, the statistical errors are much larger due to the smaller number of observed glitches~\cite{and12}. It is also worth pointing out that the analyses of the 2007 glitch detected in PSR J1119$-$6127, as well as of the 2010 glitch in PSR~2334$+$6 lead to even larger fractional moments of inertia of the crust than for Vela~\cite{alp11,akb15}. In particular, the constraint $I_{\rm s}/I\geq 0.204~\bar m_n^\star/m_n$ inferred from the glitch in PSR J1119$-$6127 cannot be fulfilled by the crustal superfluid. Indeed, inserting $I_{\rm s}\simeq 0.893 I_{\rm crust}$ and $\bar m_n^\star \simeq 5.13 m_n$~\cite{cha13} thus yields $I_{\rm crust}>I$. 

Our study suggests that the neutron superfluid in the core of a neutron star plays a more important role than previously thought. The pinning of neutron 
superfluid vortices to magnetic flux tubes opens promising perspectives for unravelling the origin of giant glitches~\cite{gug14}. This scenario could be independently 
tested by the search for gravitational waves associated with glitch events~\cite{ben10,aba11,hask15}.

\begin{acknowledgments}
The authors thank Jan Steinhoff for helpful discussions. This work was financially supported by the Fonds de la Recherche Scientifique - FNRS (Belgium) under grant
n$^\circ$~CDR J.0187.16, NSERC (Canada), the DFG within the Research Training Group 1620 ``Models of Gravity'', as well as the European Cooperation in Science and Technology (COST) Action MP1304. The hospitality of the Aspen Center for Physics (USA), where some of this work was carried out, is gratefully acknowledged. The Aspen Center 
for Physics is supported by National Science Foundation grant PHY-1066293. 
\end{acknowledgments}


\end{document}